# Unraveling Primary Energy Conversion Dynamics in Intact Cells of *Gemmatimonas groenlandica*

*Julian Lüttig[1*], Ajeet Kumar[2], Maurice Schenker[2], Alastair T. Gardiner[3], Michal Koblížek[3], Donatas Zigmantas[2*] and Jennifer P. Ogilvie[1*]*

[1] Department of Physics, University of Ottawa, 150 Louis-Pasteur Pvt, K1N 6N5 Ottawa, Canada

[2] Division of Chemical Physics, Lund University, P.O. Box 124, 221 00 Lund, Sweden

[3] Laboratory of Anoxygenic Phototrophs, Institute of Microbiology, 379 81 Třeboň, Czech Republic

ABSTRACT Photosynthetic bacteria have evolved specialized light-harvesting structures to efficiently capture sunlight. Here, we investigate the ultrafast dynamics in intact cells of *Gemmatimonas groenlandica,* featuring unique double-ring LHh-LH1 photosynthetic complexes containing bacteriochlorophyll *a*. Two-dimensional electronic spectroscopy with lock-in detection allows us to directly observe primary energy-conversion processes, from excitation-energy transfer within the antenna system to the formation of a charge-separated state in the reaction center, taking place on 50 ps time scale. Through comparison with *Rhodospirillum rubrum*, which lacks the outer antenna ring, we use a combination of global analysis and nodal/center line slope analyses to resolve LHh-LH1 excitation energy transfer which occurs on a ~1-2 ps timescale in *Gemmatimonas groenlandica*.

## Introduction

Photosynthetic bacteria have developed specific protein–pigment complexes that are highly adapted to harvest sunlight in their respective ecological niches.[1] The protein scaffold fixes the geometrical arrangement of the pigments relative to one another, tunes their local environment, thus controlling the electronic coupling between them and giving rise to delocalized excited states referred to as excitons. The photosynthetic apparatus can be divided into antenna complexes and reaction centers. In the primary steps of photosynthesis, the antenna complexes absorb sunlight and the resulting excitation energy is transported to the reaction center on an ultrafast timescale of femto- and picoseconds, followed by charge separation.[2] Charge separation is only the first step of a cascade of electron transfer processes, which result in an electrochemical gradient fueling the photosynthetic cell.

The ability to harvest light energy using (bacterio)chlorophyll-containing reaction centers is restricted to only small number of bacterial phyla. In 2014, a phototrophic bacterium, *Gemmatimonas* (*Gem.*) *phototrophica*, of the phylum Gemmatimonadota, was isolated and characterized.[3,4] A genetic analysis suggests that the photosynthesis gene cluster was acquired *via* horizontal gene transfer from purple bacteria (phylum Pseudomonadota). Therefore, it represents the first documented case when the complete set of photosynthesis genes was horizontally transferred between distant bacterial phyla. Later another phototrophic species from the same genus, *Gem. groenlandica,* was described.[5] Both species contain closely related photosynthesis gene clusters and share a similar overall organization of their photosynthetic apparatus.

The reaction-center (RC)–antenna complexes (RC-dLH) of both species, *Gem. phototrophica* and *Gem. groenlandica,* consist of two concentric antenna rings surrounding the RC.[6,7] The inner ring resembles the light-harvesting (LH) 1 complex found in purple bacteria. *Gem. phototrophica* and *Gem. groenlandica* lack the LH2 complex present in many purple bacterial species. Instead, the outer ring, which genetically originates from LH1, is optically similar to LH2. Thus, the outer ring is referred to as the LHh complex (with "h" denoting hybrid).[8] Despite this overall similarity, the antenna structures of *Gem. groenlandica* and *Gem. phototrophica* also exhibit significant differences which are most apparent in the LHh antenna complex. Due to the formation of a 3.1 Å H-bond between αTrp[31] and the α-1 BChl C3' keto group in *Gem. groenlandica* the absorption spectrum of LHh is red shifted by approximately 44 nm relative to that of LHh in *Gem. phototrophica*, leading to significant spectral overlap between LHh and LH1. In *Gem. phototrophica*, this αTrp[31] sidechain adopts a different rotamer conformation so that this H-bond is unable to form, resulting in well separated absorption bands for LH1 and LHh. Although the structures of the RC–dLH complexes in both species have been characterized, there is only scarce information about how excitation energy relaxes and is transferred between the antenna complexes and subsequently drives charge separation in the RC.

Following ultrafast exciton dynamics from antenna complexes to the RC requires femtosecond time resolution. Pump–probe spectroscopy measures transient absorption changes as a function of the delay ($t_2$), referred to as the population time, between the pump and probe pulses, but does not resolve the excitation-frequency dependence, making energy transfer between spectrally overlapping subsystems difficult to distinguish.[9] The method of choice to investigate ultrafast energy transfer is two-dimensional (2D) electronic spectroscopy.[10–14] In 2D spectroscopy, two pump pulses are used, separated by another variable delay $t_1$, referred to as the coherence time.

Fourier transformation over this delay provides the excitation frequency axis. The resulting 2D spectrum correlates excitation and detection frequencies, analogous to the correlation of nuclear-spin transitions in 2D NMR spectroscopy.[15] A 2D spectrum contains diagonal peaks and off-diagonal peaks, the latter commonly referred to as cross peaks. The amplitude change of the various peaks in 2D maps, e.g. the growth of cross peaks with population time $t_2$ reveals energy transfer between spectrally distinct states or subsystems. In addition to energy transfer, 2D spectroscopy allows one to distinguish between homogeneous and inhomogeneous broadening by lineshape analysis.[16] In photosynthetic systems featuring energy transfer between several subunits and states, 2D spectroscopy shows its full potential and has been used to reveal dynamics in reaction centers[17,18] and various light-harvesting complexes.[11,19]

Applying 2D spectroscopy, or any transmission-type ultrafast spectroscopy to study intact cells is challenging for several reasons. First, the size of photosynthetic cells is on the order of the excitation wavelength, resulting in strong scattering that can obscure the desired signals. Second, light-harvesting complexes contain densely packed and electronically coupled pigments that form extended excitonic networks. Under natural illumination, the probability of simultaneously creating multiple excitons within the same connected antenna network is negligent.[20] However, ultrashort laser pulses can create multiple excitons within the same system, which can undergo exciton–exciton annihilation if they encounter each other during their lifetime. Reducing the intensity of the pulses results in an overall lower amplitude of annihilation but also reduces the signal-to-noise ratio.[21,22] Annihilation therefore complicates the interpretation of spectroscopic signals, [23,24] particularly in intact cells with large, interconnected antenna networks and high absorption cross sections. These challenges have limited application of 2D spectroscopy to study intact cells to a handful of examples reported in literature. A 2D setup based on lock-in detection has been previously used to study the flow of energy through the photosynthetic apparatus of green sulfur bacteria at cryogenic temperature[25] and subsequently through purple bacteria *in vivo*.[26] This detection enables measurements at low pump intensities, retaining a high signal-to-noise ratio,[27] while minimizing the contribution of exciton–exciton annihilation. Another study used the Gradient Assisted Photon Echo Spectrometer, a variant of 2D spectroscopy which encodes the coherence time spatially, to measure 2D spectra of purple bacteria.[28] A recent study demonstrated 2D spectroscopy on intact cells of cyanobacteria using crossed polarizations of the pump and probe beams to minimize scattering.[29]

Here, we use a 2D setup based on lock-in detection previously applied to intact cells[25,26] and report the first *in vivo* 2D spectroscopic investigation of ultrafast excitation dynamics and charge separation in intact cells of *Gem. groenlandica*. We compare the 2D spectra of *Gem. groenlandica*, which contain both LH1 and LHh, with those of *Rhodospirillum* (*Rsp.*) *rubrum* (phylum Pseudomonadota) that contains LH1 but lacks LHh. Analysis of the lineshape evolution reveals an additional slow component in *Gem. groenlandica* that is absent in *Rsp. rubrum* and is, therefore, associated with the presence of LHh and may reflect energy transfer from LHh to LH1.

**Methods**

The 2D setup is described in detail in the literature[27] and here we only review the key features. A Yb:KGW amplified laser system (Pharos, Light Conversion Ltd.) generates 170 fs light pulses at 1030 nm and is used as a light source for pumping a lab-built non-collinear optical parametric amplifier (NOPA).[30] While the laser system can operate at repetition rates between 1 and 200 kHz, we set it to 20 kHz as a compromise between increasing the signal-to-noise ratio and minimizing contributions from long-lived species between consecutive laser shots. Double-frequency lock-in detection further suppresses signal contributions from such long-lived species to the 2D spectra. The pulses provided by the NOPA are compressed via a prism compressor to 17 fs as measured by nonlinear autocorrelation. The beam is split into two parts via a plate beamsplitter and one part is guided through a magnetic-drive delay stage controlling the population time $t_2$. The two beams then pass a transmissive diffraction grating resulting in four beams that are then focused into the sample in a non-collinear geometry.[31,32] Beams 1 and 2, which act as the pump pulses, are modulated by two choppers at 114.8 Hz and 860.9 Hz, respectively. The fourth beam acts as a local oscillator and is attenuated by a 2 OD neutral density filter. After the sample, the spectral interferogram between the local oscillator and emitted signal is measured in a spectrometer. For all measurements, the relative polarization between pulses 1 and 2, i.e., the pump pulses, and pulses 3 and 4, i.e., the probe and local-oscillator pulses, was set to the magic angle. Importantly, the sample cell was constantly moved in the xy-plane, i.e., in the plane perpendicular to the direction of the beams propagation to renew the excited sample volume, which is necessary to observe the charge-separated state of the RCs. The turnover time of the charge separated state is longer than the time between pulses. Thus, if the cuvette is not moved, the RCs accumulate in the charge-separated, or "closed," state. By continuously translating the cuvette, a small amount of fresh sample is introduced into the excitation volume between successive laser shots, ensuring that a sufficient fraction of RCs remain in the "open" state and can undergo charge separation upon photoexcitation. The range and speed of the motion need to be adjusted accordingly: if the speed is too slow signatures of "open" RCs might be too small to detect. On the other hand, with an increased range the chances of hitting scratches on the cuvette and impurities in the sample increase. For *Gem. groenlandica*, the pump pulse energy was set to 1 nJ per each pulse, while for *Rsp. rubrum*, we used 0.5 nJ per pulse.

Both intact-cell samples were dissolved in the same buffer (MES/KCl/glycerol). One major issue when working with intact cells is aggregation. To remove cell aggregates, we centrifuged the samples at low speed (a few thousand revolutions per minute) for approximately 10 min and used the supernatant for the measurements. To increase the signal-to-noise ratio, we averaged several population-time scans: seven scans for *Gem. groenlandica* and four scans for *Rsp. rubrum*. Population times were measured at 109 specific points, with increasing step size for longer population times. The coherence time was scanned from -150 to +150 fs with steps of 2 fs in the case of *Gem. groenlandica*, and from -80 to 80 fs with steps of 2 fs in the case of *Rsp. rubrum*.

**Results**

Figure 1a compares the photosynthetic units of the two species investigated in this study, highlighting the additional LHh antenna ring in *Gem. groenlandica* relative to *Rsp. rubrum*. Thus, comparing the two species allows one to identify signatures specific to the LHh complex. The absorption spectra of intact cells of *Gem. groenlandica* and *Rsp. rubrum* are shown in Fig. 1b as blue solid and red dashed lines, respectively. The absorption spectrum of *Gem. groenlandica* shows only one major broad peak at 869 nm (11507 $cm^{-1}$) without two distinct signatures of LHh and LH1. Indeed, even at 77 K, the two antenna complexes cannot be distinguished in the linear absorption spectrum and the spectral positions of the two antenna complexes can only be inferred in the 77 K linear CD spectrum.[7] A smaller signal at 800 nm (12500 $cm^{-1}$) originates from the RC. Interestingly, the absorption spectrum of *Rsp. rubrum* is almost identical in shape to that of *Gem. groenlandica*, although it is slightly red shifted, with the major absorption peak located at 890 nm (11236 $cm^{-1}$). The laser spectrum used in 2D spectroscopy experiments with an approximate

FWHM of 160 nm (Fig. 1b, pink shaded area) covers the major transition of both species equally well.

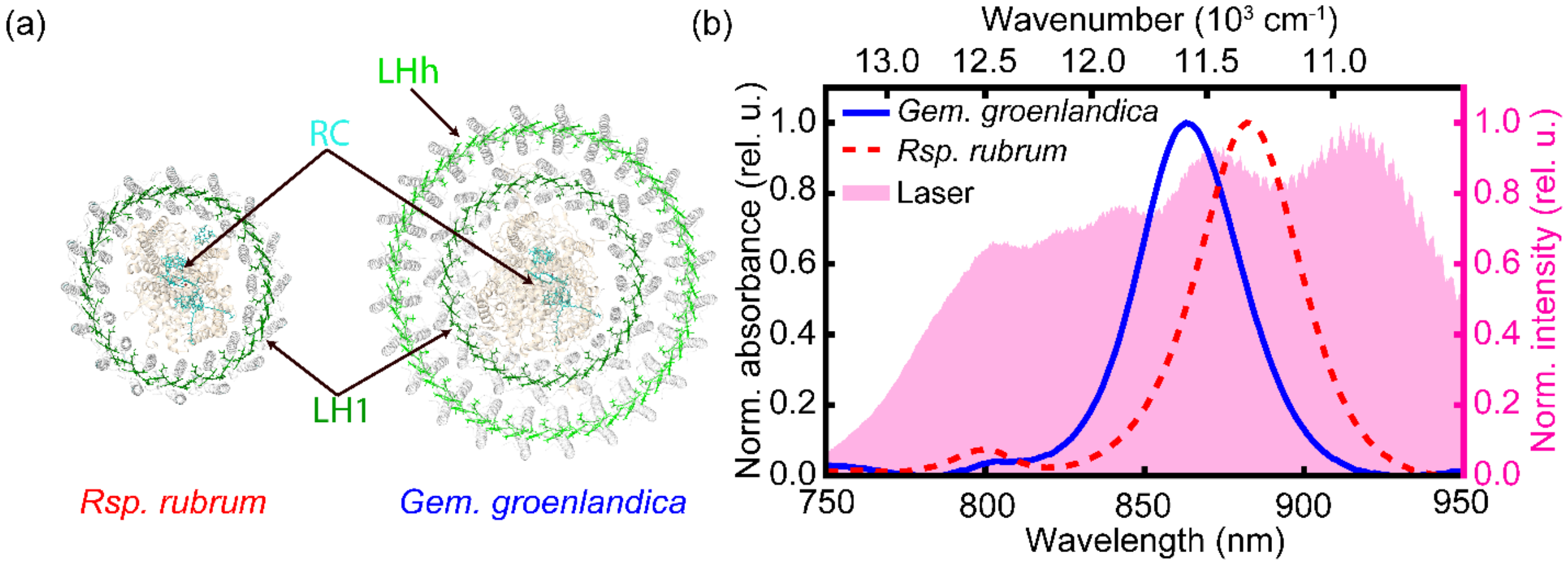

**Figure 1.** Characteristics of *Rsp. rubrum* and *Gem. groenlandica*. (a) Structures of the RC-LH1 unit of *Rsp. rubrum* (PDB: 7OY8, left)[33] and the RC-dLH unit of *Gem. groenlandica* (PDB 9H19, right).[7] LH1 is marked in dark green, the reaction center in cyan, LHh in light green, and the protein matrix in light gray. (b) Absorption spectra of intact cells of *Gem. groenlandica* (blue, solid line) and *Rsp. rubrum* (red, dashed line). The laser spectrum used in the 2D spectroscopy experiments is shown as a pink shaded area.

The 2D spectra of *Gem. groenlandica* and *Rsp. rubrum* at various population times are shown in Fig. 2 on the top and bottom rows, respectively. Each spectrum is normalized to the maximum of the spectra at $t_2 = 30$ fs, and the normalization factor is provided in the upper right corner of each spectrum. At early population times, both species exhibit one positive peak comprising ground state bleach and stimulated emission contributions and one negative peak arising from excited state absorption. The negative peak is shifted to higher detection wavenumbers. Such a lineshape is typical of large excitonic systems such as LH2,[34] LH1,[35] chlorosomes[36] or J-aggregates.[37] In both species, we observe a change in the lineshape within the first 5 ps, which is most clearly visible in the nodal line where the positive and negative signal contributions cancel.

While the nodal line is initially tilted, it becomes horizontal in both species after 5 ps. We will discuss the lineshape evolution within the first picoseconds in more detail below. Importantly, we do not observe any clear cross peaks in the 2D spectra of *Gem. groenlandica* that would point to energy transfer from LHh to LH1. However, since the absorption bands of LH1 and LHh cannot be distinguished in the linear absorption spectrum, the absence of such cross peaks is expected. The most striking difference between the 2D spectra of the two species can be found at long population times $t_2$. For *Gem. groenlandica*, the peaks observed at early population times remain visible at long population times. Additional positive and negative peaks appear at detection wavenumbers of $\tilde{\nu}_3$ =12400 $cm^{-1}$ and $\tilde{\nu}_3 = 12800$ $cm^{-1}$, respectively, both centered at $\tilde{\nu}_1 = 11600$ $cm^{-1}$. The emergence of these off-diagonal features at long population times is a typical signature of a charge-separated state within the RC and indicates that excitation energy has been transferred from the antenna complexes to the RC, followed by charge separation.[38,26] *Gem. groenlandica* contains purple bacterial type-2 RC, comprising a strongly coupled "special pair" of two BChl $a$ molecules, referred to as P, as well as two additional BChl $a$ molecules (referred to as "B") and two bacteriopheophytins (referred to as "H").[1] We assign the positive and negative features emerging at long population times and centered at $\tilde{\nu}_1 = 11600\,\mathrm{cm}^{-1}$ to the electrochromic shift of the B band. This electrochromic shift is induced by the local electric field from the charge-separated state. We note a small spectral difference of approximately $100\,\mathrm{cm}^{-1}$ on the excitation axis compared with previous studies on isolated mutant RCs from *Rhodobacter (Rba.) capsulatus* which lacked the quinones, making $P^+H_A^-$ the final charge separated state.[17] The shift may be attributed to a number of possible factors, including the lack of quinones, small geometrical differences between the different RC species, or differences in the experimental conditions, such as the laser spectrum used. Compared to other 2D spectroscopy studies on whole cells of *Rba. sphaeroides* (also known as *Cereibacter sphaeroides*) the positive and negative peak appear at the same spectral position.[26] The charge-transfer process in the RC is typically much faster than energy transfer from LH1 to P, such that excitation of P is followed rapidly by charge separation. In the case of *Rba. sphaeroides,* global analysis could not resolve these two processes separately and observed only the rate-limiting energy-transfer step from LH1 to P.[26] In *Gem. groenlandica* we expect the similar limitation in our ability to resolve the charge separation process.

In contrast to *Gem. groenlandica*, the 2D spectrum of *Rsp. rubrum* remains essentially unchanged between $t_2 = 5$ and 100 ps, and no signatures of charge separation are observed. This can be understood as a consequence of exciting the sample by a laser pulse approximately every 50 µs, an interval shorter than the RC-reopening timescale reported for *Rsp. rubrum*.[39] Translating the cuvette therefore does not fully replace the excitation volume between consecutive shots, but it gradually introduces fresh sample and so limits the buildup of closed reaction centers. Since the cuvette was translated at comparable speeds for *Gem. groenlandica* and *Rsp. rubrum*, the absence of a signature of the charge-separated state in *Rsp. rubrum* likely reflects the differences in RC-reopening kinetics between the two species. Under comparable excitation conditions, clear signatures of charge-separated states were observed earlier in *Rba. sphaeroides*.[26] Indeed, RC-reopening has been reported to be faster in *Rba. sphaeroides* than in *Rsp. rubrum*,[40] which would maintain a larger fraction of open RCs in the case of *Rba. sphaeroides* and thus enhance the charge-separated-state signatures in the 2D spectra. The reopening kinetics of *Gem. groenlandica* have not been previously reported. Since here we observe the charge-separated state in *Gem.*

*groenlandica,* we make a conclusion that its turnover (charge recombination) time is likely closer to that of *Rba. sphaeroides* than *Rsp. rubrum.*

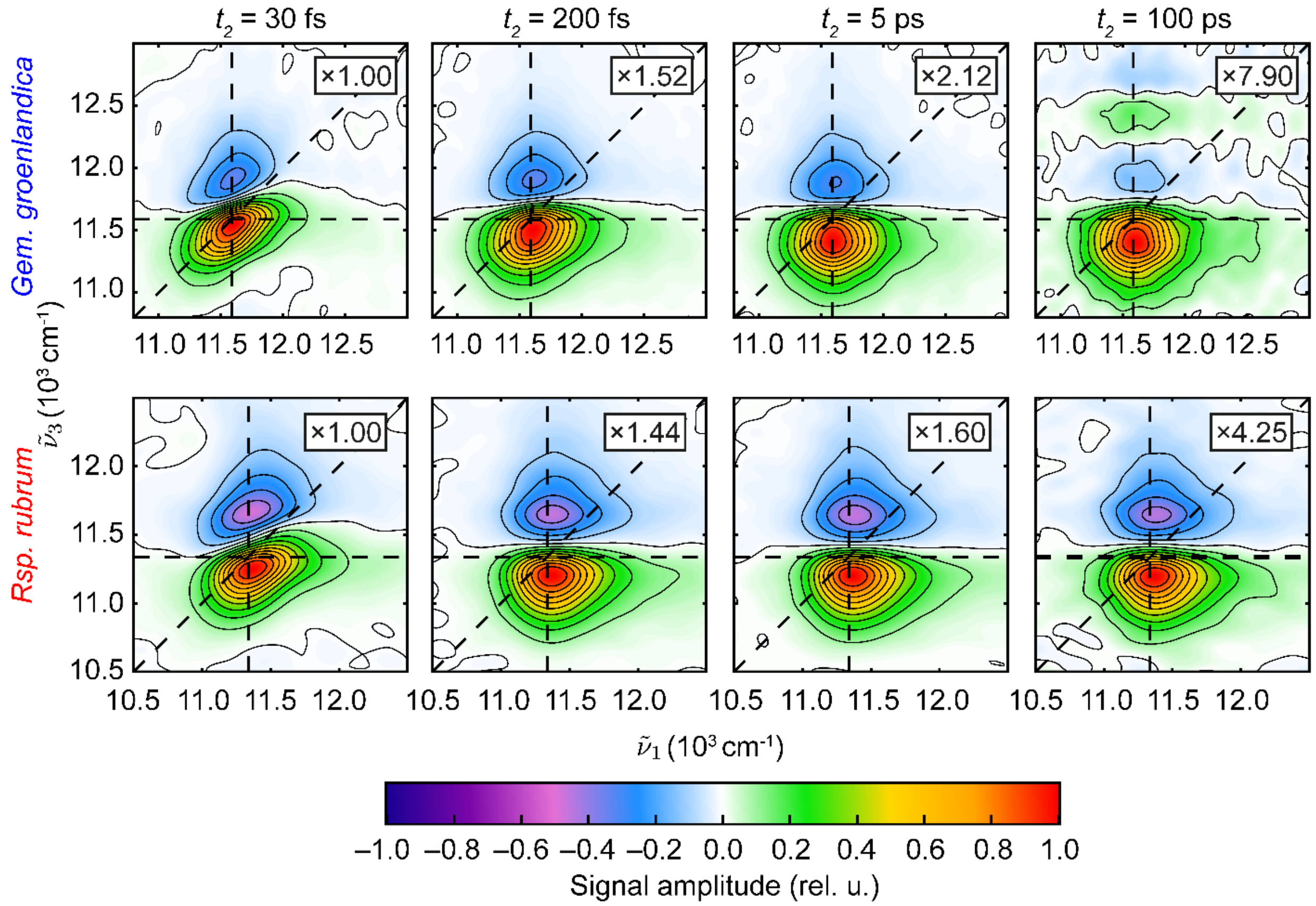

**Figure 2.** Population time dynamics of *Gem. groenlandica* (top row) and *Rsp. rubrum* (bottom row). Each 2D spectrum is normalized to the maximum of the corresponding spectrum at $t_2 = 30$ fs. The normalization factor for each 2D spectrum is shown in the upper-right corner. Dashed lines are drawn at the diagonal and the peak maximum of the absorption spectra at 863 nm (11588 cm$^{-1}$) and 882 nm (11338 cm$^{-1}$) for *Gem. groenlandica* and *Rsp. rubrum*, respectively. Contour lines are drawn from $-1$ to $+1$ with a spacing of 0.1.

We applied global analysis to the 2D spectra, a standard method for separating spectral components with distinct dynamics.[17,25,41,42] In global analysis the complete 2D dataset is fitted with a sum of exponential components and a connected 2D spectral amplitude which represent the decay associated spectra (DAS). When the underlying dynamics follow exponential decays and the corresponding time scales are sufficiently separated, the DAS can be used to identify specific processes such as energy transfer. In total, we could reliably separate five contributions for *Gem. groenlandica* which are depicted in Fig. 3. The first component, $\tau_1$, shows a positive feature along the diagonal, tilted compared to the main spectral features at later times. The shape and the 52 fs timescale together suggest that this component reflects ultrafast spectral diffusion and dynamic localization.[43] Furthermore, pulse overlap effects, typically summarized as coherent artifacts, may also contribute to the first DAS component.[44,45] The 810 fs component, $\tau_2$, shows a positive peak above the horizontal dashed line at $\tilde{\nu}_3 = 11600$ cm$^{-1}$ and a negative peak below it. These features are opposite in sign to those observed in the 2D spectrum on this timescale. However, both peaks in the DAS are shifted towards higher excitation wavenumbers relative to the corresponding features in the 2D spectra, consistent with energy transfer from higher-lying to lower-lying states on a timescale of 810 fs. We can thus assign this component to energy transfer from LHh to LH1. The DAS associated with the 3.8 ps component resembles the 2D spectrum at $t_2 = 5$ ps, corresponding to an overall decay of the positive and negative peaks. Such an overall decay on a picosecond timescale is typical of exciton–exciton annihilation.[25,26,28] The DAS associated with the 53 ps component shows positive and negative peaks near the diagonal, as well as an additional negative peak at $(\tilde{\nu}_1, \tilde{\nu}_3) = (11600, 12400)\,\mathrm{cm}^{-1}$ and a weak positive peak at $(\tilde{\nu}_1, \tilde{\nu}_3) = (11600, 12800)\,\mathrm{cm}^{-1}$. At the latter two positions, the 2D spectra for different population times show a positive and negative peak assigned to the spectroscopic signatures of the electrochromic shift of the B band and the corresponding peaks in the DAS reflect how this specific spectral signature of the electrochromic shift is formed over time. Accordingly, we assign the $\tau_4$ component to the formation of the charge-separated state, following energy transfer from the antenna complexes to the RC and ensuing charge separation. Finally, the last component with $\tau_5 > 1$ ns resembles the same shape as the 2D spectrum at $t_2 = 100$ ps. The corresponding DAS therefore represents the decay of the charge-separated state, and the lifetime of this state exceeds the experimentally accessible population-time range. Note, that the positive peak is shifted to lower wavenumbers along the detection axis reflecting ground-state bleach and stimulated emission contributions from P rather than contributions from LH1.

The global analysis of *Rsp rubrum* is shown in Section S1 of the Supporting Information. In contrast to the global analysis of *Gem. groenlandica*, we can only distinguish three different components: a fast component of 52 fs corresponding to ultrafast spectral diffusion and pulse overlap, a 16 ps component corresponding to exciton–exciton annihilation, and a 141 ps component corresponding to the overall lifetime decay. The absence of the 810 fs component in *Rsp. rubrum* supports its association with LHh-related dynamics in *Gem. groenlandica*. Additionally, no charge-separated state is detected in *Rsp. rubrum*, consistent with the absence of the 53 ps component, and longer charge recombination time as described above. Global analysis can successfully separate different components such as energy transfer and annihilation components in this case. However, exciton–exciton annihilation does not necessarily follow a single exponential decay and might, therefore, "leak" into other components especially when the time scales are close to each other.[26] The standard mitigation strategy to deal with exciton–exciton annihilation is to reduce the excitation intensity and thus operate in a regime in which the third-order nonlinear signal dominates, i.e., when mostly single-particle dynamics are probed.[20,22,46] However, with lower pump intensity the signal-to-noise ratio is also decreased, and it becomes difficult to quantify the remaining amount of annihilation. Following this strategy, we performed additional measurements at a lower pump pulse energy (Section S2 of the Supporting Information). However, the results were inconclusive, as it was difficult to disentangle annihilation contribution in global analysis of 2D spectra (Section S2 of the Supporting Information). At lower pump powers, another problem specific to this system becomes apparent: the relaxation and annihilation components, associated with timescales of 810 fs and 3.8 ps, respectively, have partially overlapping kinetics and similar spectra, and are therefore difficult to distinguish even by global analysis. We therefore decided to follow a second strategy, lineshape analysis, to provide a more accurate estimate of the energy transfer dynamics from LHh to LH1.

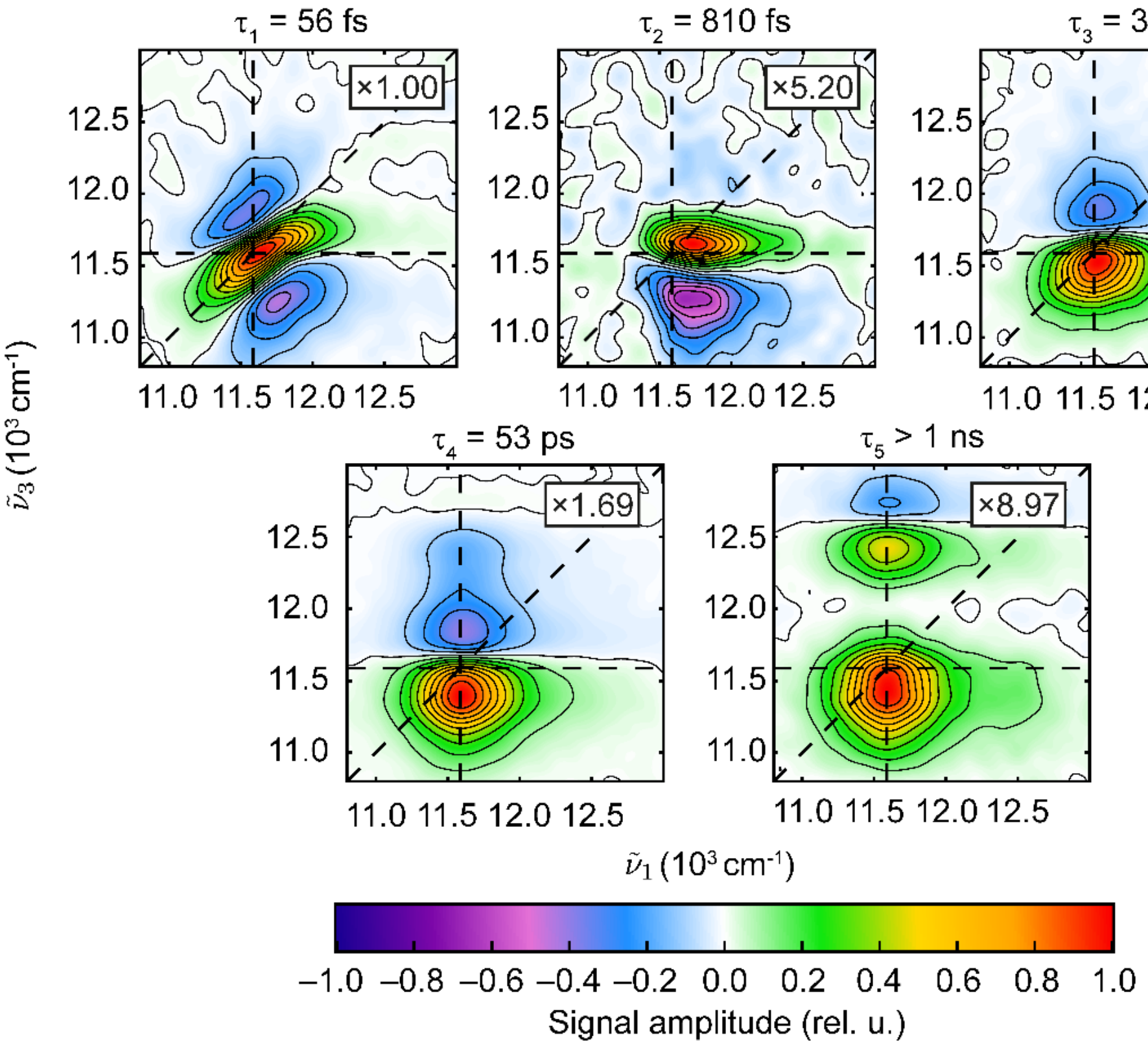

**Figure 3.** Decay associated spectra of *Gem. groenlandica*. All 2D spectra are normalized to the maximum of the $\tau_1$ DAS component and the corresponding normalization factor is given in the top right corner of each map.

An advantage of 2D spectroscopy is that the nonlinear response signal is distributed over two frequency axes allowing to analyze lineshapes in detail.[43,47,48] Different strategies have been employed in the past to detect and interpret changes in the lineshape, such as the center line slope analysis or nodal line analysis.[49–51] Such analyses can be used to detect signatures of the interaction with the environment, i.e., spectral diffusion. A comprehensive analysis of center line slope dynamics in electronic 2D spectroscopy was provided by Šanda and coworkers using the example of zinc phthalocyanine,[51] where oscillations in the center line slope dynamics were attributed to vibrations. Other studies focused on the ultrafast dynamics of the Stokes shift by analyzing slices along the excitation axis of the 2D spectrum, revealing oscillations and ultrafast solvation.[47] Anna et al. analyzed the lineshapes of the 2D spectra of the PSI trimers of *Thermosynechococcus elongatus* to probe fast energy transfer.[52] They observed that lineshape changed from diagonally elongated to a more circular shape on a 50 fs timescale, suggesting ultrafast relaxation of the $Q_{y(0\text{–}0)}$ transition. Here, we combine nodal line slope and center line slope analyses to find signatures specific to LHh. We note that the lineshape of the 2D spectra depends on the phasing procedure in which the absolute phase of the 2D signal is determined typically by comparing the integrated 2D spectrum with a transient absorption spectrum based on the projection slice theorem.[32,53] For both types of cells, we compared the integrated 2D spectra with a broadband pump–probe measurement for selected population times to determine the absolute phase. Since *Rsp. rubrum* lacks LHh but contains LH1, it provides an appropriate reference for identifying spectral signatures associated with LHh.

The nodal lines in the 2D spectra of *Gem. groenlandica* for population times of $t_2 = 50$ fs and $t_2 = 300$ fs are shown in Fig. 4a. Clearly, at later population times, the nodal line is not horizontal, i.e., not parallel to the $\tilde{\nu}_1$ axis. In contrast, the nodal line of *Rsp. rubrum,* while tilted for early population times, is completely horizontal for later population times (Fig. 4b). The change of the nodal line slope with population time $t_2$ for *Gem. groenlandica* and *Rsp. rubrum* is shown in Fig. 4c and d, respectively. For easier comparison, we plot the lineshape dynamics of both species on the same time scale. In both cases, the nodal line slope changes rapidly within the first 100 fs. A second, slower component can be observed in the case of *Gem. Groenlandica,* which is notably absent in *Rsp. rubrum.* The slow component is also present for the low-power measurement (Section S2 of the Supporting Information). Accordingly, the nodal line slope of *Gem. groenlandica* can be described by a biexponential decay with time constants of $T_1 = 65 \pm 6$ fs and $T_2 = 1600 \pm 400$ fs. Since the second, slow component is absent in the nodal line slope dynamics of *Rsp. rubrum*, the data can be fitted with a single-exponential decay with a time constant of $T_1 = 62 \pm 4$ fs. The center line slope analysis yields similar results and timescales, as shown in Fig. 4e for *Gem. groenlandica* and Fig. 4f for *Rsp. rubrum*, consistent with the nodal line slope analysis discussed above. Plots with an expanded early-time axis for *Rsp. rubrum* are shown in Section S3 of the Supporting Information. We note that biexponential dynamics of the center line slope were also found in the case of chlorophyll *a* in methanol with time scales of 500 fs and 6.9 ps and the corresponding dynamics were interpreted in the context of solvation.[43] We expect that solvation processes in *Gem. groenlandica* and *Rsp. rubrum* are similar, since we used the same buffer and experimental conditions for both systems. Given that *Rsp. rubrum* contains only LH1 and lacks LHh, we attribute the additional slow component observed in both the nodal line slope and center line slope dynamics in *Gem. groenlandica* to effects connected to LHh.

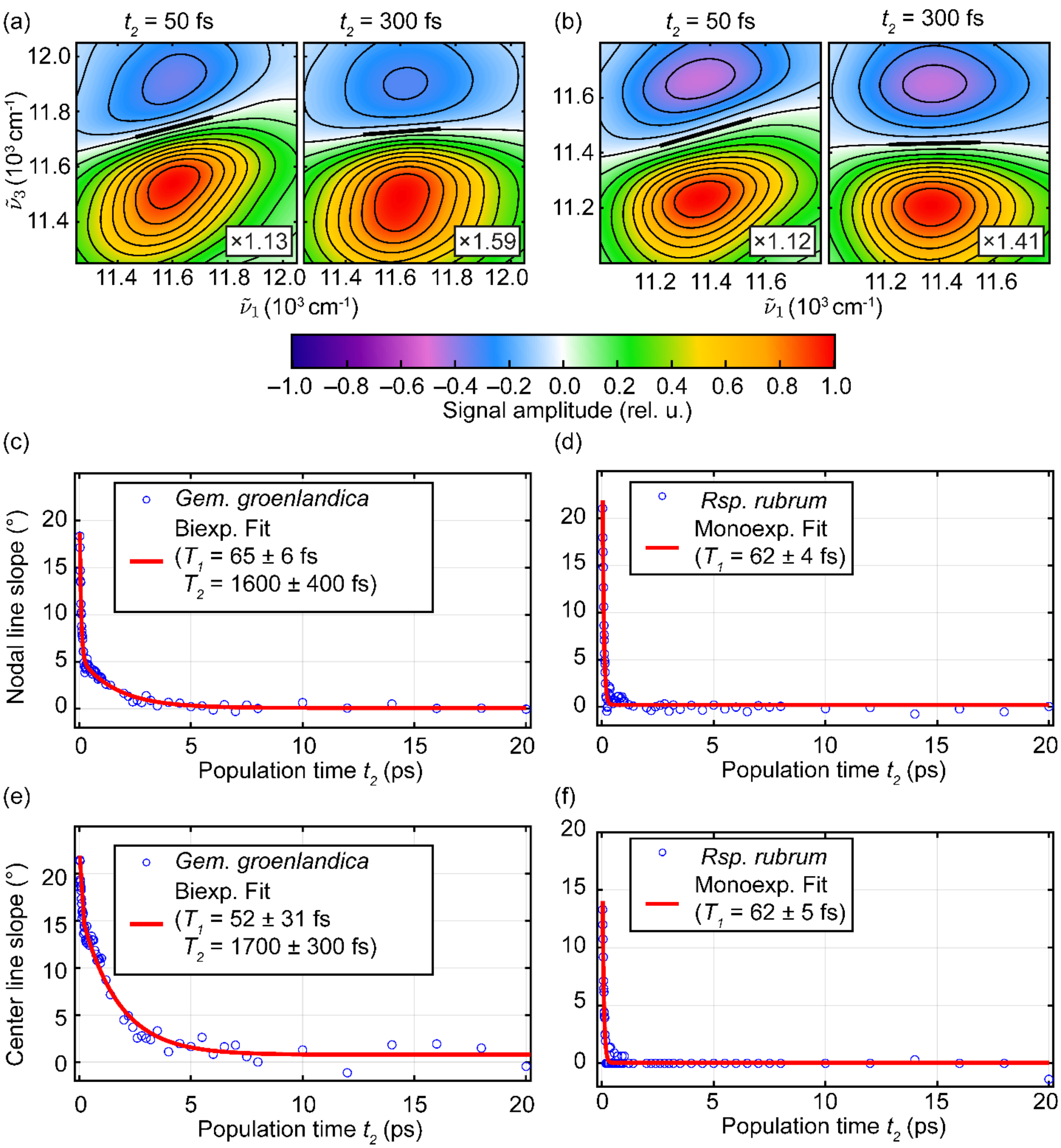

**Figure 4.** Nodal line slope and center line slope analyses of *Gem. groenlandica* and *Rsp. rubrum*. 2D spectra at $t_2 = 50$ fs and $t_2 = 300$ fs for (a) *Gem. groenlandica* and (b) *Rsp. rubrum.* The nodal lines are shown in black, and the normalization factors relative to the spectrum at $t_2 = 30$ fs are given in the lower right corner. Nodal line slope over the population time $t_2$ for (c) *Gem.*

*groenlandica* (blue circles) and (d) *Rsp. rubrum* (blue circles). Biexponential fits are shown as red solid lines. Centre line slope over the population time $t_2$ for (e) *Gem. groenlandica* (blue circles) and (f) *Rsp. rubrum* (blue circles). A biexponential fit and monoexponential fit are shown as red solid lines for *Gem. groenlandica* and *Rsp. rubrum,* respectively.

In general, two processes may contribute to the additional nodal line slope and center line slope dynamics in *Gem. groenlandica*: energy transfer from LHh to LH1 and the internal relaxation within the excitonic manifold of LHh. In our analysis we cannot distinguish between the two effects. The two extracted time components of the biexponential fit of the line slope dynamics might tentatively be connected to fast internal relaxation and slower energy transfer from one antenna complex to another. Energy transfer between antenna subsystems typically occurs on timescales ranging from hundreds of femtoseconds to several picoseconds and would therefore be consistent with the slower component of the biexponential fit. In *Gem. phototrophica* energy transfer from LHh to LH1 was reported to be 2 ps.[6] Other systems show similar slow energy transfer components. Comparable picosecond timescales have been reported for energy transfer between LH2 and LH1 in *Rba. sphaeroides* at room temperature, with characteristic timescales of approximately 3–5 ps and 8 ps in our recent study of intact cells.[26] Both methods, global analysis and lineshape analysis, reveal a component associated with the presence of LHh. Global analysis yields a timescale of approximately 810 fs, whereas nodal line slope and center line slope analyses yield somewhat longer timescales of approximately 1.6 ps. These values are of the same order of magnitude, although both approaches, lineshape analysis and global analysis, have sources of error. Exciton–exciton annihilation may influence the time constant extracted by global analysis because its multiexponential dynamics are difficult to separate from other processes. Annihilation might also affect the lineshape analyses, and its contribution may be different in *Rsp. rubrum* and *Gem. groenlandica*. However, our lineshape analyses of *Gem. groenlandica* show very similar dynamics at half the pulse energy, indicating that the influence of annihilation may be minor (see SI Fig. S4). Additionally, we cannot exclude intracomplex relaxation within the LHh antenna that might influence the timescales observed in the lineshape analyses. However, given similar architecture of the photosynthetic apparatus in *Gem. groenlandica* and *Gem. phototrophica*[7], the similar respective LHh to LH1 transfer times of 1.6 ps and 2 ps seem reasonable. We expect that future simulations based on the available crystal structures will shed light on the microscopic processes underlying the observed dynamics.

In summary, we performed the first 2D experiments on *Gem. Groenlandica in vivo,* and used *Rsp. rubrum* as a reference system to identify dynamics associated with LHh. Using a lock-in-detection 2D spectroscopy setup, we suppressed strong scattering and achieved a high signal-to-noise ratio at low pump intensities, thereby minimizing contributions from exciton–exciton annihilation. We detected clear signatures of ultrafast relaxation and the formation of the charge-separated state in *Gem. groenlandica*. Global analysis further resolved the formation of the signature of the charge-separated state on a timescale of 50 ps. Direct comparison with *Rsp. rubrum*, which lacks LHh, allowed us to identify dynamics associated with the unique LHh antenna of *Gem. groenlandica*. Global analysis revealed a component with a timescale of approximately 810 fs, absent in the measurements of *Rsp. rubrum*, while nodal line slope and center line slope analyses provided a corresponding component of 1.6 ps. These components were absent in *Rsp. rubrum* and are therefore associated with the presence of LHh. The ~1-2ps timescale is consistent with LHh-to-LH1 energy transfer in similar photosynthetic bacteria, although contributions from intracomplex relaxation within LHh cannot be excluded. Upcoming simulations and extended experiments will explore these processes in greater detail. and determine their respective rates.

ASSOCIATED CONTENT

**Supporting Information**. Results of global analysis of *Rsp. rubrum,* Low-pulse energy measurement of *Gem. groenlandica* and corresponding lineshape analysis, Lineshape dynamics of *Rsp. rubrum*

AUTHOR INFORMATION

**Corresponding Author**

*jluettig@uottawa.ca

*jogilvi3@uottawa.ca

*donatas.zigmantas@chem.lu.se

**Author Contributions**

J.L. and J.P.O. planned the research. A.T.G. and M.K. prepared and characterized the bacterial samples. J.L., A.K., and M.S. conducted the 2D experiments under the supervision of D.Z., and J.L. analyzed the data and wrote the first draft of the manuscript together with J.P.O. All authors reviewed and revised the manuscript and approved the final version.

**Notes**

The authors declare no competing financial interests.

ACKNOWLEDGMENT

J. L. acknowledges support from the Human Frontier Science Program (Grant No. LT0056/2024-C). A. T. G. and M. K. were funded by the Photomachines project (CZ.02.01.01/00/22_008/0004624) financed by the OP JAK program of the Czech Ministry of Education. J. P. O. acknowledges support from the Natural Sciences and Engineering Research

Council of Canada (NSERC) (grant RGPIN-2025-06687). D. Z., A. K. and M. S. acknowledge the support of the Swedish Research Council and the Carl-Trygger Foundation.